\documentclass[prx,twocolumn,amsmath,amssymb,showpacs,superscriptaddress,groupedaddress]{revtex4}
\usepackage{graphicx}
\usepackage{dcolumn}
\usepackage{bm}
\usepackage[FIGTOPCAP]{subfigure}
\usepackage[usenames]{color}
\usepackage[all]{xy}
\def \bea{\begin{eqnarray}}
\def \eea{\end{eqnarray}}

\begin{document}

\title{The active trap model}
\author{Eric Woillez}
\affiliation{Department of Physics,
Technion-Israel Institute of Technology,
Haifa, 3200003, Israel.}
\author{Yariv Kafri}
\affiliation{Department of Physics,
Technion-Israel Institute of Technology,
Haifa, 3200003, Israel.}
\author{Nir S. Gov}\affiliation{Department of Chemical and Biological Physics, Weizmann Institute of Science, Rehovot 7610001, Israel.}

\begin{abstract}
Motivated by the dynamics of particles embedded in active gels, both in-vitro and inside the cytoskeleton of living cells, we study an active generalization of the classical trap model. We demonstrate that activity leads to dramatic modifications in the diffusion compared to the thermal case: the mean square displacement becomes sub-diffusive, spreading as a power-law in time, when the trap depth distribution is a Gaussian and is slower than any power-law when it is drawn from an exponential distribution. The results are derived for a simple, exactly solvable, case of harmonic traps. We then argue that the results are robust for more realistic trap shapes when the activity is strong.
\end{abstract}


\maketitle

\emph{Introduction.} In-vitro experiments have probed the
non-thermal (active) fluctuations in an "active gel", which is most often realized as a
network composed of cross-linked actin filaments and myosin-II molecular motors \cite{backouche2006active,Schmidt,marina,Mackintosh}. The fluctuations
inside the active gel are measured using the tracking of tracer particles, and was used to demonstrate the non-equilibrium nature of these systems through the breaking of the Fluctuation-Dissipation theorem (FDT) \cite{Mackintosh}. In these active gels, myosin-II molecular motors generate relative
motion between the actin filaments, through consumption of ATP, and thus drive the athermal random motion of the probe particles dispersed throughout the network. Similar motion of tracer particles was observed in living cells \cite{fredberg,fodor2015activity}.

In both the in-vitro gels, and in cells, over short times, the tracer particle seems to perform caged random motion, while trapped in the elastic network. On longer times it is observed
that the actin network allows the tracer to perform "hopping" diffusion, as it makes large amplitude motions \cite{Schmidt,fredberg,fodor2015activity,sonn2017dynamics,sonn2017scale}, driven by the same active forces. This large scale motion was treated on a coarse-grained scale in \cite{fodor2015activity} \footnote{Note that in many experiments the molecular motors produce forces that on very long times break-up the network, which is coarsening, due to the internal stresses \cite{backouche2006active,marina,e2014time}. This prevents the analysis of very long-time dynamics in these systems.}.

Here we explore in more detail the process by which active forces can drive hopping diffusion in a heterogeneous medium. We use a trap model \cite{bouchaud1992weak} where the particle is assumed to be trapped in a potential well of variable depth, representing the structural inhomogeneity present in the system. The particle is affected by random active forces, which eventually "kick" the particle from the well. This event can correspond to the release of the tracer particle from the confining network, or more generally to the triggering of some unspecified rearrangement of the constituents of the system. After each such event, the particle (system) is locked in a new confining organization, and a new activated escape process begins. 

The distribution of potential well depths determines the type of hopping diffusion performed by the particle. Indeed, it is well known that for a thermal system, a Gaussian distribution of potential depths gives rise to normal hopping diffusion, while an exponential distribution of potential depths can give rise to anomalous diffusion: $\langle x^2\rangle\propto t^\alpha$, $0<\alpha<1$. (for a review see  \cite{Bouchaud1990}). This result is a direct consequence of the Kramers escape rate which is exponential in the depth of the trap. For active systems the picture can be different \cite{angelani2014first,geiseler2016kramers,sharma2017escape}. Indeed, recently it was shown for a class of active particles \cite{woillez2019activated,dhar2019run} that the
escape time from a trap depends on the detailed structure of the potential \cite{woillez2019activated}. Importantly, it is not a simply exponential function of the potential depth. By studying a new class of escape problems appropriate for active gels we find that the non-trivial behavior of the escape rate leads to several surprising features. Specifically, we find
that anomalous diffusion can appear even for a Gaussian distribution of potential depths. Furthermore, an exponential distribution of potential depths gives rise to a {\it super-slow diffusion} $\langle x^2\rangle\sim e^{C\sqrt{\ln t}}$ where $C$ is a positive constant. This result is particularly relevant, since experiments indicate that the thermal motion of tracer particles in bio-polymer gels is well described by a hopping model, with anomalous (sub-) diffusion, which therefore suggest that the distribution of traps is exponential \cite{wong2004anomalous}.

\emph{Active trap model.}
We consider a particle in a one dimensional trap described by a potential $U(x)$ and
kicked randomly by thermal and active forces \cite{ben2015modeling}. The
Langevin equation for the particle's velocity $v$ (in a simplified
one dimension reaction coordinate, with the mass
set to one) is
\begin{eqnarray}\label{eq:active process}
\dot{v}&=&-\lambda v +f_a+f_T-\frac{\partial U(x)}{\partial x}
\label{vdot}
\end{eqnarray}
where $\lambda$ is the effective friction coefficient. The thermal force $f_T$ is an
uncorrelated Gaussian white noise: $\langle f_T(t)f_T(t')
\rangle=2\lambda T_B\delta(t-t')$, with $T_B$ the ambient
temperature, and Boltzmann's constant is set to $k_B=1$.
The active force $f_a$ arises from the independent action of
$N_m$ molecular motors, each motor producing pulses of average force
$f_0$ for a duration $\tau'_p$ (either a constant or drawn from
a Poissonian process with an average value $\tau'_p$, i.e. shot-noise). This sets the persistence time of the active forces. In the large $N_m$ limit the time-evolution of $f_a$ can be approximated by the Ornstein-Ulhenbeck process
\begin{equation}\label{eq:OU process}
    \dot{f}_a=-\frac{1}{\tau'_p}f_a+\sqrt{\frac{2D}{\tau'_p}}\xi(t),
\end{equation}
where $\xi(t)$ is a standard Gaussian white noise, and
$D=N_m\langle f_0^2\rangle$. The amplitude of the active force in this model has a Gaussian distribution, which naturally arises in a spatially extended system that has a uniform distribution of active force sources, even if the individual sources (such as molecular motors) produce discrete forces (the individual forces are integrated in such extended systems \cite{alvarado2017force}). Note that the thermal force, $f_T$ is typically much smaller than the active force and can therefore be ignored in most cases. It should also be noticed that the amplitude of the white noise vanishes in Eq.~(\ref{eq:OU process}) when $\tau'_p\rightarrow0$ and the model described therefore differs somewhat from the well known active Ornstein Uhlenbeck process (the AOUP model, see \cite{fodor2016far} and references therein), even if the results for both models can be easily related.

Note that the AOUP process can be viewed as the limit of a particle that is affected by a large number of spatially distributed sources of active forces (such as molecular motors). These combine to give a force distribution that can be well approximated by a Gaussian form with a single relaxation time~\cite{ben2015modeling,DanGolan2019}. We expect our results to apply to more complex temporal correlations (\cite{sevilla2019generalized} for example) when on long-time scales a single relaxation time  dominates. However, a more careful examination of such correlations remains for future studies.

To analyze the trap model we need to evaluate the escape rate of the particle from the trapping potential. As stated above we first do this for the exactly solvable case of a harmonic potential and then turn to comment on more general potentials.

\emph{Escape from a harmonic trap.}
Here we take $U(x)=kx^2/2$ and postulate that the particle escapes from the trap when it reaches $x=a$. $k$ is expected to be proportional to the bulk modulus of the gel which depends on the gel density, cross-linker density and stiffness of the network filament. It is well known that the steady-state distribution of active particles positions, when the potential is unbounded, is very different from that of an equilibrium case  \cite{hanggi1995colored,szamel2014self,das2018confined,solon2015active,pototsky2012active,ben2015modeling,razin2019signatures,malakar2019exact,DanGolan2019}. Indeed, as we show below the escape time from the trap also behaves very differently.

The active escape from a single harmonic well of the model described above was extensively explored (using numerical simulations and analytic results in certain limits) in \cite{DanGolan2019}. In fact, the linearity of Eq. (\ref{eq:active process}) and (\ref{eq:OU process}) allows for an analytic solution in the limit of deep potential wells and small diffusion, which is detailed in the appendix. One finds that the escape time is given by a Kramer's like form:
\begin{equation}
\label{eq:underdamped_escape}
    \tau'_{\rm esc} = \tau'_0(k)e^{ka^2/2T_{\rm eff}}
\end{equation}
where the "effective temperature" $T_{\rm eff}$, assumed to be small compared to the potential height, is given by the mean potential energy
\begin{equation}\label{Teff}
   T_{\rm{eff}}=k\langle x^2 \rangle = \frac{D (1+ \lambda \tau'_p)}{\lambda(\lambda + k \tau'_p +{\tau'_{p}}^{-1})} \;.
\end{equation}
Importantly, $T_{\rm{eff}}$ depends on the shape of the potential \footnote{Note that we use the concept of effective temperature because of the mathematical analogy between the mean escape time Eq. (\ref{eq:underdamped_escape}) and the classical Arrhenius law, but $T_{\rm{eff}}$ does not correspond here to any thermodynamical concept.}. This implies, in stark contrast to the equilibrium problem where the escape time grows exponentially in $k$, that it grows here exponentially in $k^2$ for large $k$. This, as we show below has important implications for the generalized active trap model. We note that the validity of Eqs. (\ref{eq:underdamped_escape}) and (\ref{Teff}) was numerically tested in \cite{DanGolan2019}, over a wide range of parameters.

In fact, as one might expect, the same phenomenology also appears in the overdamped limit which can be readily solved. The overdamped limit is obtained by taking the limit of large friction and neglecting the inertial term in Eq. (\ref{eq:active process}). Moreover, for simplicity, we assume that $f_T\ll f_a$ and neglect thermal noise. Rescaling time $t'=\frac{t}{\lambda}$, and $\tau_p=\tau'_p/\lambda$, Eqs. (\ref{eq:active process})-(\ref{eq:OU process}) become
\begin{equation}
\begin{cases}
    \dot{x}=-\frac{\partial U(x)}{\partial x} +f_a,\label{eq:AOUP_process}\\
    \dot{f}_a=-\frac{1}{\tau_p}f_a+\sqrt{\frac{2D}{\tau_p}}\xi(t) \;.
    \end{cases}
\end{equation}
Note the model is mathematically equivalent to the standard Active Ornstein-Uhlenbeck Particles (AOUPs) studied, for example, in \cite{koumakis2014directed,marconi2015towards,szamel2015glassy,fodor2016far,caprini2018active}, but with a noise amplitude which scales differently with $\tau_p$. For a harmonic potential one can introduce the standard change of variable \cite{caprini2019active} $p=-kx+f_a$ to obtain
\begin{equation}\label{eq:equilibrium map}
\begin{cases}
    \dot{x}= p,\nonumber\\
    \dot{p} =-\gamma p -\nabla U_{\rm{eff}}+\sqrt{\frac{2\gamma D}{1+\tau_p k}}\xi(t).
    \end {cases}
\end{equation}
These equations describe an equilibrium Brownian particle in a potential $U_{\rm{eff}}=\frac{kx^2}{2\tau_p}$, experiencing a friction $\gamma=\frac{1}{\tau_p}+k$, and temperature $T_{\rm{eff}}=\frac{D}{1+\tau_p k}$. The change of variables maps the problem to an equilibrium one. The standard Kramers law then follows immediately with the mean escape time from a trap of size $a$, when the effective temperature is small, given by
\begin{equation}\label{eq:active_arrhenius}
    \tau_{esc}= \tau_0(k) e^{\frac{U_{\rm{eff}}(a)}{T_{\rm{eff}}}}= \tau_0(k) e^{\frac{ka^2(1+\tau_p k)}{2D\tau_p}},
\end{equation}
where $\tau_0(k)$ is a subexponential correction. Again $T_{\rm{eff}}$ is an explicit function of the potential through $k$. We note that the steady-state distribution in a harmonic potential was found previously, using a different method, in \cite{szamel2014self}. Curiously, we find that the persistence time $\tau_p$ does not affect the long-time hopping diffusion (dominated by large $k$, Eq.\eqref{eq:active_arrhenius}), although it does affect $T_{\rm{eff}}$ \cite{DanGolan2019}.

The main result of Eq. \eqref{eq:active_arrhenius}, similar to Eq. \eqref{eq:underdamped_escape}, is that the mean escape time is exponential in $k^2$. This is in stark contrast to the standard Arrhenius law for passive Brownian particles which gives an escape time which in exponential in $k$.

We now demonstrate that this new scaling dramatically affects the dynamics of an active particle in a network of random traps. We consider a standard trap model \cite{bouchaud1992weak}, where the particle once escaping from the harmonic well through an active process falls into a new trap whose depth is drawn from a random distribution. For simplicity we initially assume that the traps are of equal size $a$ so that at each hop between two traps the particle is displaced by a distance $a$. The long time behavior of the system is dominated by the distribution of traps of large depth (equivalently of large stiffness $k$), so we may reduce Eq.\ref{eq:active_arrhenius} to
\begin{equation}\label{eq:simple_arrhenius}
    \ln(\tau_{esc})\approx \frac{k^2a^2}{2D}.
\end{equation}
where additional terms do not change qualitatively the results. We consider two canonical choices for the distribution of trap depths, a Gaussian and an exponential.

\emph{Gaussian distribution of trap stiffness:} We first consider a normal distribution of trapping potential depths, encoded in the local network stiffness $k$,
\begin{equation}
P(k) \propto e^{[-(ka^2-k_0a^2)^2/\sigma_E^2]} \;. \label{gaus}
\end{equation}
with $k_0$ the stiffness at the distribution peak, and $\sigma_E$ the stiffness variance. For thermal activation it is straightforward to check that this leads to a Log-normal distribution of trapping times. Therefore the mean escape time is finite and the motion of the particle in the network is diffusive with a mean-square displacement growing as $t$. In contrast, for the active escape, using Eq.\eqref{eq:simple_arrhenius}, we obtain in the limit of large $\tau_{esc}$ that the distribution of trapping times is given by
\begin{eqnarray}
P(\tau_{\text{esc}})=P(k)\left|\frac{\partial k}{\partial\tau_{\text{esc}}}\right| \sim \frac{1}{\tau_{\text{esc}}^{1+\mu}\sqrt{\ln(\tau_{\text{esc}})}},
\label{eq:lognormalactive}
\end{eqnarray}
with $\mu=\frac{2Da^2}{\sigma_E^2}$.
This distribution is normalizable, but has a diverging first moment when $\mu<1$, which leads to anomalous diffusion \cite{bouchaud1992weak}. For what follows, it is useful to recall how this anomalous diffusion can be obtained from a simple scaling argument \cite{denisov2010continuous} (we comment that another, similar approach, is to use the framework of continuous time random walk, see for example \cite{barkai2000continuous,barkai2003aging,klafter1987stochastic}).

Consider the total time $T_N$ needed for the particle to perform $N$ step. Assuming that each step takes a time $\tau_i$ drawn from the distribution (\ref{eq:lognormalactive}), and that the random variables $\tau_i$ are independent, we have $ T_N=\sum_{i=1}^N \tau_i$. When $\mu<1$, the distribution Eq.(\ref{eq:lognormalactive}) has no average and the total time $T_N$ is dominated by the maximal escape time $\tau^*$. This value can be estimated using $N\int_{\tau^*}^\infty P(\tau_{\text{esc}})d\tau_{\text{esc}} \approx 1$.
With the distribution of escape times Eq. (\ref{eq:lognormalactive}) we get $N\sim (\tau^*)^\mu$, which leads to $T_N\sim N^{\frac{1}{\mu}}$.
Finally, using $\langle x^2\rangle\sim Na^2$ one finds that for $\mu<1$, the distribution of escape times (\ref{eq:lognormalactive}) gives the anomalous behavior
\begin{equation}\label{eq:anomalous Gaussian}
   \frac{ \langle x^2(t)\rangle}{a^2}\sim t^\mu.
\end{equation}
For $\mu>1$ the argument above implies that motion becomes diffusive with $\langle x^2(t)\rangle/a^2\sim t$.
For acto-myosin gels the limit of anomalous diffusion corresponds to low active forces, large stiffness and large stiffness variance. Note that the dependence on the persistence time $\tau_p$ cancels out in the ratio that determines the anomalous regime.

\emph{Exponential distribution of trap stiffness:} Next, let us consider an exponential distribution of energy barriers: $P(\Delta E)\sim \exp{[-\Delta E/E_0]}$. For such a distribution the thermal escape-time distribution is a power-law \cite{bouchaud1992weak}, which gives a diverging first moment, with anomalous behavior similar to the Gaussian active case described above, for $T/E_0<1$. In passive bio-polymer (actin) gels, thermal diffusion of tracer particles was observed to fit this description of anomalous sub-diffusion \cite{wong2004anomalous}, suggestive of an exponential trap distribution. We now show that the behavior for the active case is very different.

For the active hopping, using arguments similar to the Gaussian case, the exponential distribution of barriers heights gives that the distribution of escape times for large times is given by
\begin{equation}
P(\tau_{\text{esc}})\sim \frac{1}{\tau_{\text{esc}}\sqrt{\ln(\tau_{\text{esc}})}}e^{-\frac{a\sqrt{2D}}{2E_0}\sqrt{\ln(\tau_{\text{esc}})}}.
\label{expactive}
\end{equation}
The above distribution has a diverging mean. Using exactly the same arguments as for the Gaussian distribution of stiffness with $\mu<1$,
we get a diffusion process that for long times grows more slowly than any power-law with
\begin{equation}
\frac{\langle x^2(t)\rangle}{a^2}\propto e^{\frac{a\sqrt{2D}}{2E_0}\sqrt{\ln(t)}} \;.
\label{eq: anomalous exponential}
\end{equation}
This behavior falls in the class of \emph{superslow} diffusion process \cite{denisov2010continuous}, where $\langle x^2(t)\rangle/t^\alpha\to0$ as $t\to\infty$, for all $\alpha>0$. In particular, in the limit of $t\to\infty$ the slope of the mean square displacement approaches zero.

Finally, we note that similar to the case of normal distribution, we find that the super-slow diffusion is independent of the persistence time $\tau_p$, in Eq. (\ref{eq: anomalous exponential}).

\emph{Coupled trap-size and stiffness distribution in a biopolymer gel:} So far we considered traps of finite spatial extent $a$, and an independent distribution of stiffness values. In a real biopolymer gel, such as cross-linked actin filaments, the fluctuations in the local stiffness are often determined by fluctuations in the local concentration of cross-linkers and lengths of the biopolymer filaments \cite{gardel2004elastic,kasza2010actin}. In these systems the local stiffness $k$ and the mesh pore size $a$ depends on the cross-linkers density
\begin{eqnarray}
k\propto\rho_c^{\alpha} ,\quad a\propto \rho_c^{-1/3} \Rightarrow \Delta E\propto\rho_c^{\alpha-2/3}
\end{eqnarray}
where $\alpha$ was found to vary between $1$ and $2$ \cite{gardel2004elastic,kasza2010actin}. The energy barriers are therefore expected to depend on the cross-linker's concentration as a power-law in the range: $\Delta E\propto\rho_c^{1/3}$,$\rho_c^{4/3}$.

Obtaining detailed distributions of pore sizes and cross-linker's concentration from the experimental data is difficult. Recent measurements of the pore-size distribution in an acto-myosin gel found it to be similar to a Gaussian distribution \cite{ideses2018spontaneous}. However such gels are highly heterogeneous, evolve over time \cite{e2011active}, and are predicted theoretically to approach a critical state with power-law distribution of its structural heterogeneity \cite{alvarado2013molecular}.

\emph{Extension to general potentials:}
Our results have been obtained assuming harmonic traps. One might wonder how general these results are for more general trap shapes. In \cite{DanGolan2019} it was shown that the escape rate from a single potential well does not depend strongly on the shape, for monotonously increasing potentials and a Gaussian active noise (Eqs. (\ref{eq:AOUP_process})). We now discuss in the following the mean escape time from a general trap the two limits of the over-damped model, $\tau_p\ll \tau_{U}$ and $\tau_U\ll\tau_p$, where $\tau_U$ is the typical relaxation time inside the trap, and is given by the order of magnitude of $\frac{\partial^2 U(x)}{\partial x^2}$.

The limit of small correlation time $\tau_p\ll \tau_U$ has been thoroughly studied in the past \cite{hanggi1995colored,marconi2016velocity}. It has been shown that Eqs. (\ref{eq:AOUP_process}) reduce to the equilibrium escape problem for a passive particle with an effective temperature $T_{\rm eff}=D\tau_p$. The mean escape time in this limit is thus given by the classical Arrhenius law with an "effective temperature". The first non-equilibrium correction, for a potential whose derivative at the escape point is zero, comes at order $\tau_p$ and depends on the full shape of the trap \cite{hanggi1995colored}. One recovers to leading  order the results of the classical thermal trap model \cite{bouchaud1992weak}, with $T_{\rm eff}=D\tau_p$. In particular, the mean escape time in this limit only depends on the depth of the trap and not on its shape.

We next consider the interesting regime where the time $\tau_p$, related to binding and unbinding a molecular motor, is still much smaller that the time to escape using a thermal fluctuation, but much larger than the relaxation time $\tau_U$ inside a trap. We thus still neglect thermal noise and take the limit $\tau_p\gg \tau_U$. We consider  Eqs. (\ref{eq:AOUP_process}) with $t'=t/\tau_p$
\begin{equation}\label{active nonequilibrium}
    \begin{cases}
    \dot{x}=\tau_p\left(f_a-\frac{\partial U(x)}{\partial x}\right), \\
    \dot{f}_a=-f_a+\sqrt{2D}\xi(t).
    \end{cases}
\end{equation}
In the limit $\tau_p\gg \tau_U$, the position of the particle has to satisfy $f_a=\frac{\partial U(x)}{\partial x}$. At every time, the particle is in a local equilibrium state where the active force balances the potential force. For a general barrier one expects $\frac{\partial U(x)}{\partial x}$ to have a single maximum located at a point $x_{cr}$ where $\frac{\partial^2 U}{\partial x^2}(x_{cr})=0$. Defining the maximal force that the potential can exert as $F_{\rm max}=\max\{\frac{\partial U}{\partial x}\}$, the particle needs a force fluctuation that reaches at least $F_{\rm max}$ to escape from the trap. For the stochastic process (\ref{eq:OU process}), such a fluctuation occurs within a typical time
\begin{equation}
\label{eq:OU_fluc}
    \tau_{\rm ou} \approx \tau_p e^{ \frac{\left(F_{\rm max}\right)^2}{2D}}.
\end{equation}
Once the particle has crossed the critical point, the force $f_a-\frac{\partial U(x)}{\partial x}$ is strictly positive, and the particle moves quickly to the escape point, according to the first equation in (\ref{active nonequilibrium}). The mean escape time for a general trap, in the limit $\tau_p \gg\tau_U$, is given by Eq. (\ref{eq:OU_fluc}).
The limit of large correlation time much larger than the thermal escape time from the trap has been studied in \cite{DanGolan2019} and leads to a different result.

Eq. (\ref{eq:OU_fluc}) illustrates the fact that in the large correlation time limit an active particle is sensitive to the maximal force it experiences, whereas passive particles are sensitive to the depth of a trap. This dependence on the shape of the potential seems to be rather generic for active particles \cite{woillez2019activated}. As there is no linear dependence between the maximal force exerted by a potential and its depth, all equilibrium results relying on the classical Arrhenius law fail for active particle. For the particular case of an harmonic trap presented in this paper, $F_{\rm max}=ka$, and we recover the result (\ref{eq:simple_arrhenius}) using Eq. (\ref{eq:OU_fluc}). Importantly, one expects the depth of the potential to be generically correlated to its maximal slope. This suggests that the results describe above are rather robust. Note that even if this correlation leads to an escape time of the form $\ln(\tau_{esc})\approx A{\Delta E^s}$, with the above results suggesting $s \geq 2$, $A$ a constant, and $\Delta E$ the trap depth, the super-slow diffusion results remains almost unchanged for an exponential distribution of $\Delta E$, behaving as $\langle x^2(t)\rangle\propto e^{B(\ln(t))^{1/s}}$, with $B$ a constant. Furthermore, for a Gaussian distribution of $\Delta E$ one obtains a super-slow diffusion with $P(\tau_{\rm esc}) \sim  e^{-C (\ln \tau_{\rm esc})^{2/s}}$, with $C$ a model dependent constant.

\emph{Discussion.}
Our key result is that active escape gives rise to anomalous hopping diffusion even in systems that have rather uniform energy landscapes. For the case of energy barriers with a normal distribution, we predict a regime of anomalous (sub-)diffusion. For exponential distribution of energy barriers, which can occur naturally in different self-assembled systems, we predict an even more extreme behavior: super-slow diffusion. These results arise from the quadratic relation between the logarithm of the mean escape time and the depth of the energy barrier, unlike the linear relation obtained for thermal activation. Note that in experimental realizations of active gels the cross-linkers are dynamic and the system tends to exhibit viscous flow on very long time-scales \cite{backouche2006active,marina,e2014time}. This may prevent the observation of the long-time regime where the anomalous or super-slow diffusion appears. However, since the thermal diffusion was found to be anomalous in biopolymer gels \cite{wong2004anomalous}, future studies of the motion of tracer particles over long times in active acto-myosin gels \cite{sonn2017scale}, could probe the super-slow diffusion regime that we predict. Note that often the active sources, such as molecular motors, are themselves affected by the elastic restoring forces of the trap \cite{razin2019signatures}, and the consequences of this on the escape process remain to be explored.

We have neglected thermal noise in our analysis, both for simplicity and since in many active systems it is much smaller than the active component (see for example \cite{sonn2017dynamics,sonn2017scale}). In the appendix we include thermal noise in the calculation, and show explicitly that at very large values of trap depth the escape time is controlled by the thermal process. This gives a cut-off time above which the escape process is dominated by the thermal noise and the standard behaviors of the classical trap model are recovered. These time scales are, however, very long and most likely very difficult to observe.

The basic property of the active force that gives rise to this different activation dynamics is its persistence time. This means that any process of activated diffusion, rearrangement, or flow that is driven by forces with a finite correlation (persistence) time, will give rise to the anomalous dynamics we described. Dynamics that are driven by such non-thermal forces appear in many non-equilibrium systems. Moreover, since Gaussian and exponential distributions of energy barriers are common in many disordered systems, we find that active (non-thermal) motion over such an energy landscape can very easily become anomalous, and exhibit aging behavior. These results should be relevant to dynamics inside living cells, artificial active matter and driven disordered systems. In systems with active processes that have many time scales, our results should apply as long as only a single one is dominant in the long-time limit.

\begin{acknowledgments}
We thank Golan Bel for discussions that initiated this study, and useful comments. This work is made possible through the historic generosity of the Perlman family. NSG is the incumbent of the Lee
and William Abramowitz Professorial Chair of Biophysics and this research was supported by the Israel Science Foundation
(Grant No. 580/12). EW \& YK acknowledge support from the Israel Science Foundation and an NSF5-BSF grant. EW is supported by a Technion fellowship.
\end{acknowledgments}

\bibliography{activeGel_EscapeBib}
\begin{widetext}
\appendix
\section{Derivation of the mean escape time for the underdamped model of active
particles}

In the following, we consider an active particle in a harmonic well
of infinite depth $U(x)=\frac{1}{2}kx^{2}$
\begin{eqnarray}
\dot{x} & = & v\nonumber \\
\dot{v} & = & -\lambda v-kx+f_{a}\nonumber \\
\dot{f}_{a} & = & -\frac{1}{\tau_{p}}f_{a}+\sqrt{\frac{2D}{\tau_{p}}}\xi(t),\label{eq:active AOUP}
\end{eqnarray}
where $\xi(t)$ is the standard Gaussian white noise, $f$ is the
active force with correlation time $\tau$, and $\lambda$ is the
friction, and $D=\left\langle f_{0}^{2}\right\rangle $. Eq. (\ref{eq:active AOUP})
corresponds to Eqs. (1-2) of the main text. The stationary Fokker-Planck
equation for the process (\ref{eq:active AOUP}) is 
\begin{equation}
-\partial_{x}\left(vP_{s}\right)-\partial_{v}\left(\left(-\lambda v-kx+f_{a}\right)P_{s}\right)-\partial_{f_a}\left(-\frac{1}{\tau_{p}}f_{a}P_{s}\right)+\frac{D}{\tau_{p}}\partial_{f_{a}}^{2}P_{s}=0.\label{eq:FP}
\end{equation}
 As the dynamics is linear, the stationnary distribution $P_{s}$
is Gaussian. We thus solve the Fokker-Planck equation using the ansatz
\begin{equation}
P_{s}=Ce^{-\frac{1}{2}X^{T}AX},\label{eq:ansatz}
\end{equation}
where $X=(x,v,f_{a})$ is the vector of all coordinates, and $A$
is a $3\times3$ symmetric matrix. Using (\ref{eq:ansatz}) in Eq.
(\ref{eq:FP}), the Fokker-planck equation turns into a linear system
for the coefficients of $A$ that can be solved easily analytically.
We find
\[
A=\frac{\tau_{p}}{D}\begin{bmatrix}k\lambda\left(k+\frac{1}{\tau_{p}^{2}}\right) & k\lambda^{2} & k\lambda\\
k\lambda^{2} & \lambda\left(\frac{1}{\tau_{p}^{2}}+\frac{2\lambda}{\tau_{p}}+k\lambda+\lambda^{2}\right) & \lambda\left(\lambda+\frac{1}{\tau_{p}}\right)\\
k\lambda & \lambda\left(\lambda+\frac{1}{\tau_{p}}\right) & \frac{1}{\tau_{p}}+\lambda
\end{bmatrix}.
\]
The space probability distribution $\rho_{s}(x)$ can be obtained
by summing the invariant measure over $v$ and $f_{a}$
\[
\rho_{s}(x)=\int\int P_{s}(x,v,f_{a})~{\rm d}v~{\rm d}f_{a}.
\]
The final result is
\begin{equation}
\rho_{s}(x)=C_{x}\exp\left(-\frac{kx^{2}}{2}\frac{\lambda\left(\lambda+k\tau_{p}+\frac{1}{\tau_{p}}\right)}{D\left(1+\lambda\tau_{p}\right)}\right),\label{eq:final measure}
\end{equation}
with $C_{x}$ given by the normalization constrain $\int\rho_{s}(x){\rm d}x=1$
\[
C_{x}=\sqrt{\frac{k\lambda\left(\lambda+k\tau_{p}+\frac{1}{\tau_{p}}\right)}{2\pi D\left(1+\lambda\tau_{p}\right)}}.
\]
This allows us to identify the effective temperature of the main text
\[
T_{\rm eff} = \frac{D(1+\lambda \tau_p)}{\lambda (\lambda+k\tau_p+\frac{1}{\tau_p})}
\]

Let us now turn to the problem of the mean escape time from a harmonic
trap. We consider a harmonic potential $U(x)$ truncated
at $|x|=a$, so that the particle escapes the trap if it
reaches $x=a$ or $x=-a$. We now explain how the mean escape time can
be obtained from Eq. (\ref{eq:final measure}) asymptotically in the
small $D$ limit. The methods used are standard, and we review them for completeness. In principle, the mean escape time $\left\langle T_{esc}\right\rangle $
can be computed from the Fokker-Planck equation with absorbing boundary
conditions
\[
\left\langle T_{esc}\right\rangle =\int_{0}^{+\infty}tJ(t){\rm d}t.
\]
where  $J(t)$ is the outgoing flux at $x=\pm a$.
 This is in general a difficult problem. However, in the small
$D$ limit in Eq. (\ref{eq:active AOUP}), the problem simplifies. In this limit, the outgoing
flux is exponentially small in $1/D$, such that the system relaxes quickly
to a quasistationary state given by
\begin{equation}
P_{QS}(x,t)\sim\rho_{s}(x)e^{-\frac{t}{\left\langle T_{esc}\right\rangle }},\label{eq:pqs}
\end{equation}
where $\rho_{s}$ is given by Eq. (\ref{eq:final measure}) up to exponentially small corrections in $1/D$. This
relation amounts to saying that the escape is a Poisson process with
rate $\frac{1}{\left\langle T_{esc}\right\rangle }$. 

Moreover, the outgoing flux $J(t)$ is given by the probability $P_{QS}(a,t)$
to be on the boundary at time $t$, times the velocity $V_{f}$ of
the fluctuation path leading to the boundary
\begin{equation}
J(t)\sim V_{f}\rho_{s}(a)e^{-\frac{t}{\left\langle T_{esc}\right\rangle }}.\label{eq:jqs}
\end{equation}
On the other hand, we have the standard relation
\[
J(t)=-\frac{{\rm d}}{{\rm d}t}\int_{-a}^{a}P_{QS}(x,t){\rm d}x.
\]
Using Eq. (\ref{eq:pqs}), we get the relation
\[
J(t)=\frac{1}{\left\langle T_{esc}\right\rangle }e^{-\frac{t}{\left\langle T_{esc}\right\rangle }},
\]
and combining with Eq. (\ref{eq:jqs}) gives
\[
\left\langle T_{esc}\right\rangle ^{-1}\sim V_{f}\rho_{s}(a).
\]
We obtain an effective "Arrhenius law"
\begin{equation}
\left\langle T_{esc}\right\rangle \sim V_{f}\sqrt{\frac{2\pi D\left(1+\lambda\tau_{p}\right)}{k\lambda\left(\lambda+k\tau_{p}+\frac{1}{\tau_{p}}\right)}}e^{\frac{ka^{2}}{2}\frac{\lambda\left(\lambda+k\tau_{p}+\frac{1}{\tau_{p}}\right)}{D\left(1+\lambda\tau_{p}\right)}}\label{eq:arrhenius}
\end{equation}
 We strongly emphasize that the last relation is valid only in the
small $D$ limit, which corresponds here to the inequality $D\ll k^2a^2$. If this latter  inequality is not satisfied, the quasistationnary approximation Eq. (\ref{eq:pqs}) fails and subexponential corrections become more and more important in
Eq. (\ref{eq:arrhenius}). In Eq. (\ref{eq:arrhenius}),
the leading behavior for small $D$ is given by the exponential term, and one usually
gets rid of the prefactors with the approximation
\[
\left\langle T_{esc}\right\rangle \underset{}{\approx}e^{\frac{ka^{2}}{2}\frac{\lambda\left(\lambda+k\tau_{p}+\frac{1}{\tau_{p}}\right)}{D\left(1+\lambda\tau_{p}\right)}}.
\]
The exponential factor has a non-linear dependance in the trap depth
and accounts alone for the non-equilibrium behavior of the model described
by Eqs. (\ref{eq:active AOUP}).

Finally, we note that it is straightforward to extend the above results to include a thermal noise of amplitude $2 \lambda^2 D_T$ in the equation for the velocity. The resulting steady-state probability distribution takes the form
\begin{equation}
\rho_{s}(x)=C^T_{x}\exp\left(-\frac{kx^{2}}{2{\bf T_{\rm eff}}}\right),\label{eq:final measure}
\end{equation}
with a modified effective temperature ${\bf T}_{\rm eff}$ given by
\[
{\bf T}_{\rm eff}=\frac{D(1+\lambda \tau_p)}{\lambda (\lambda+k\tau_p+\frac{1}{\tau_p})}+T
\]
with $T=D_T/\lambda$ the thermal temperature and $C^T_x$ a new normalization given by 
\[
C^T_{x}=\sqrt{\frac{k}{2\pi {\bf T}_{\rm eff}}}.
\]
This illustrates that the thermal temperature can control the probability distribution, and therefore the escape rate, but only when
\[
T \gg \frac{D(1+\lambda \tau_p)}{\lambda (\lambda+k\tau_p+\frac{1}{\tau_p})} \;.
\]
\\
This only happen for extremely stiff traps where $kT \gg D(1+\lambda \tau_p)/\lambda \tau_p$. For active systems one typically has that the magnitude of the active fluctuations, quantified by $\tau_p D$, is a few orders of magnitude larger than $T$ the thermal temperature. Therefore, the results presented are expected to hold over a long-range of time before a cutoff introduced by the thermal fluctuations leads to a cross over to the standard trap model.

\end{widetext}
\end{document}